# Bulk Magnesium Diboride – Mechanical and Superconducting Properties

Vitali F. Nesterenko
Department of Mechanical and Aerospace Engineering
UCSD Materials Science and Engineering Program
University of California, San Diego
La Jolla, California 92093-0411

## Introduction

Superconductivity in $MgB_2$ was recently discovered when a porous stoichiometric mixture of Mg and B powder was sintered at 700° C for 10 hours under a high argon pressure, 196 MPa, using a hot isostatic pressing (HIP) furnace [1]. Because no encapsulation step was reported this approach apparently could not result in dense material. Magnesium diboride is a brittle intermetalllic compound which start to decompose at relatively low temperatures, about 1000° C. To overcome the problem with magnesium volatility powder-in-tube (PIT) method was successfully applied for sintering of magnesium diboride and fabrication of iron-clad $MgB_2$ ribbons [2]. But PIT method or similar methods of sintering do not result in a fully dense material. At the same time there is a need of developing methods to fabricate fully dense $MgB_2$ to enhance superconducting properties, mainly $J_c$, $H_{c2}$, and $H_{irr}$, to increase environmental robustness, mechanical strength and fracture toughness. Also porosity is detrimental for achieving a high, mirror-like quality of surface, specifically important for some applications like microwave devices and for processing methods like laser ablation and magnetron sputtering. Fundamental research on properties of magnesium diboride also needs fully dense material.

To fabricate such material several authors employed a 12-mm cubic multi-anvil-type press at a pressure of 3000 MPa and a temperature of 950° C to synthesize samples with diameters of ~4.5 mm and heights of ~3.3 mm [3-6]. Unfortunately this approach does not allow scaling of sample sizes above few millimeters and can not be used for samples of complex shapes.

Indrakanti, Nesterenko et al., [7,8], Frederick, Maple et al., [9] fabricated fully dense magnesium diboride at significantly smaller pressures using HIPing cycle with cooling under pressure (DMCUP). Complex of superconducting and other properties of HIPed material was reported in [10-13]. This paper is focused mainly on the results obtained using HIPing in comparison with samples sintered under high pressure.

## Hot Isostatic Pressing of Magnesium Diboride

Magnesium diboride powders of –325 mesh size and 98% purity were obtained from Alfa Aesar, Inc. Powder was placed into rubber jacket and cold isostatically pressed at a pressure about 5 kbar to density 65% of theoretical value. After this step, the sample was covered with a tantalum foil, in some runs zirconium foil was additionally used as a getter. The whole assembly was sealed in a pyrex capsule under a vacuum of $10^{-2}$ torr with pressure transmitting media between sample and capsule (Figure 1).

HIPing of encapsulated $MgB_2$ powder is carried out in an ABB Mini-HIPer using a DMCUP (dense material cooled under pressure) cycle where the cooling of the densified material was performed under pressure. This is helpful to reduce a local residual tensile stresses connected with temperature gradients and thermal mismatch between components. Conditions on independently controlled temperature and pressure that resulted in the best properties of bulk samples are depicted in Figure 2. Information on other HIPing conditions and corresponding results can be found in [7,8].

Cylindrical sample released from pressure transmitting media is shown in Figure 3. Ta foil becomes very brittle and can be easily separated from the dense sample after the HIPing. Pressure transmitted media was designed in a way to make sample release without excessive application of force to prevent it from fracture and to avoid chemically assisted removal of capsule. This is especially important for small sizes of samples of for complex shapes.

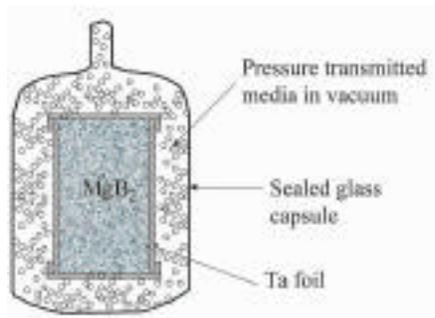

**Figure 1.** Cold isostatically pressed magnesium diboride wrapped in Ta foil and encapsulated in vacuum in glass capsule for HIPing.

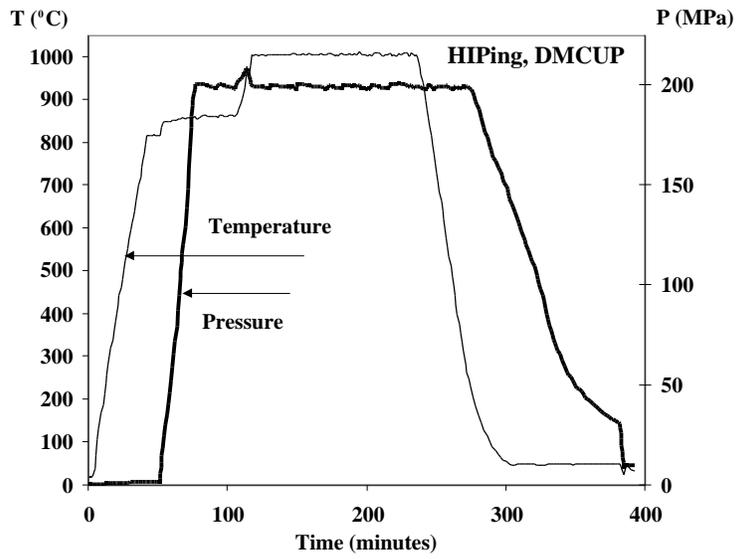

**Figure 2.** DMCUP HIP cycle employed to fabricate dense $MgB_2$ [7,8].

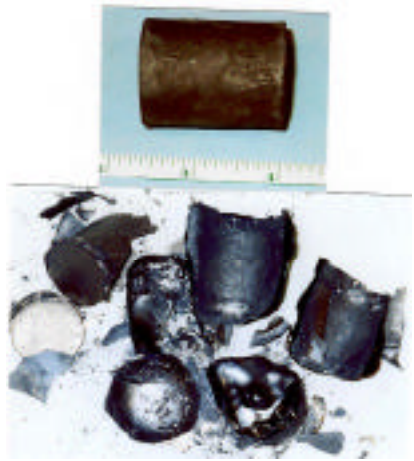

**Figure 3.** Cylindrical sample of magnesium diboride (top, scale in inches) released from pressure transmitted media (remnants are shown on the bottom).

## Mechanical Properties
### Density

The density of the synthesized material using the DMCUP HIPing cycle at 1000° C was 2.666 g/cm$^3$ (sample 1, diameter 15 mm) and appeared to be higher than the reported theoretical density based on X-ray measurements (2.625 g/cm$^3$, [14]) and similar to the density reported in [5] after sintering at 3 GPa. Measurements were made based on an ASTM B328 standard. The experimental error ($\pm 0.004$ g/cm$^3$) was found based on the comparison of the density measurements of pure aluminum samples (wire, 99.999% Al) with the known density of solid aluminum, 2.6989 g/cm$^3$ [15]. At the same time, cylindrical sample II hot isostatically pressed at the same conditions (Figure 2), but with diameter 30 mm exhibits a slightly lower density 2.56 g/cm$^3$, corresponding to the porosity volume $f$ about 4%. It is probable that optimized mechanical and superconducting properties of brittle magnesium diboride will employ some controlled level of porosity to enhance fracture toughness and introduce flux pinning centers. That is why it is important to investigate properties of MgB$_2$ with low porosity.

### Elastic constants

Elastic constants of magnesium diboride are important from fundamental point of view because they determine phonon spectrum responsible for superconducting Cooper's electron pairs. It is also important for design and fabrication of composite superconductors where metal components, presumably metal wires or tubes are load bearing and magnesium diboride supports superconducting current.

There is a lack of experimental information on elastic constants of MgB$_2$ [16]. It is mainly due to the fact that most methods (like PIT) results in porous product and samples obtained using high pressure are small for ultrasonic measurements (but certainly enough for resonant ultrasound spectroscopy (RUS), [17]). Recent quantum-mechanical calculations based on density functional theory [16] gave value for bulk modulus equal 139.3 GPa being consistent with some theoretical predictions using different approaches (140.1 GPa [18]) and significantly different with other theoretical results (163 GPa [19] and 150 GPa [20]). Data on X-ray powder diffraction in diamond anvil at $P$ 8 GPa gave value of $K = 120$ GPa [21] and 151 GPa [22].

The quality of samples after hot isostatic pressing allowed preparation of samples with high accuracy of dimensions and with sharp corners suitable for resonant ultrasound spectroscopy [17]. Four samples were tested using RUS method. Sample I-1 had dimensions 5.112*4.128*2.737 mm, sample II-1

(9.510*8.004*4.692 mm), sample II-2 (6.994*4.888*9.231 mm) and sample II-3 (8.347*6.464*4.703 mm). Faces of parallelepiped were parallel within accuracy 5 micron except one side of sample II-2 (corresponding to length) where accuracy was inside 20 microns. Geometry and quality of samples is crucial for this technique. A sample in the form of parallelepiped slightly held on corners between two transducers is shown in Figure 4.

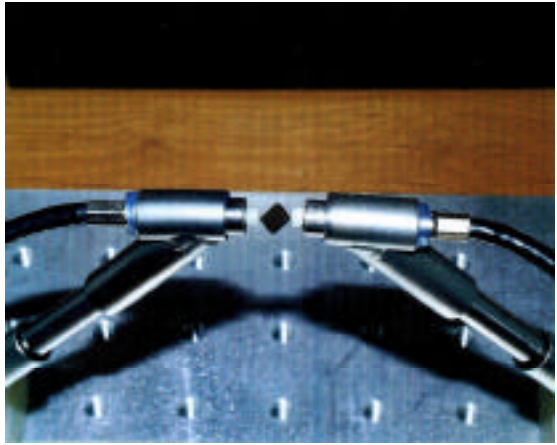

**Figure 4**. A sample of magnesium diboride (sample I-1) is slightly hold by opposite corners between transducers in RUS method. One of transducers record amplitude of vibrations for sample driven through multiple resonances by other transducer [17].

In this technique the elastic properties are characterized based on the measurements of the large number of resonant frequencies on one sample with exceptionally high absolute accuracy [17]. Elastic spectrum for samples taken from two different HIPing runs (sample I and samples II-1 – II-3 are taken from densified cylinders with diameters 15 mm and 30 mm correspondingly, in both cases the same HIPing cycle (Fig.2) was employed). The vibrational spectrum for samples I-1 and II-1 are shown in Fig. 5.

Measured elastic properties $c_{11}$ and $c_{44}$ (with the error bars, root means squared, rms) and with number of peaks used to obtain these values for all four samples are shown in Table 1.

Values of Young's ($E$) and bulk ($K$) modulus and Poisson ratio ($v$).were calculated based on elasticity theory [23] and are also shown in Table 1.

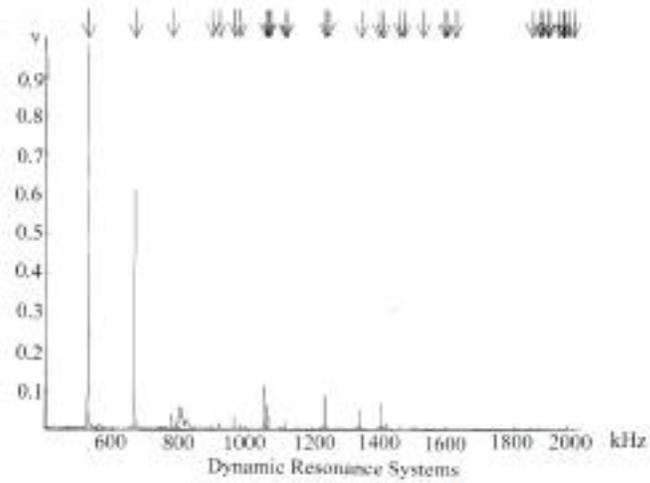

(a)

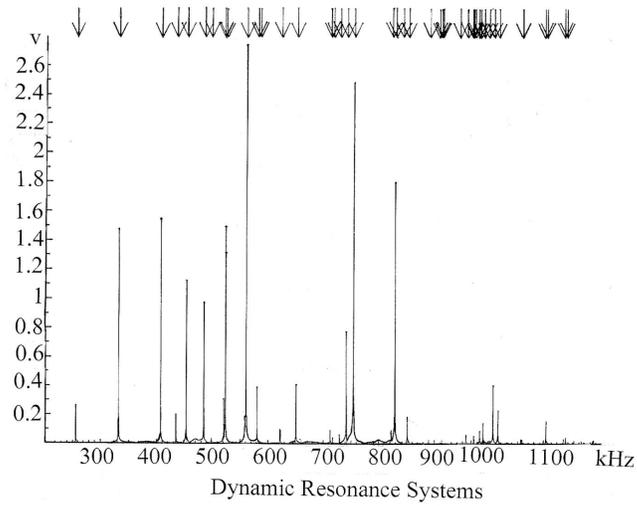

(b)

**Figure 5**. RUS vibrational spectrum for sample I-1 (a) and sample II-1 (b).

**Table 1**. Measured and calculated elastic properties of dense $MgB_2$.

|  | $C_{11}$, GPa | $C_{44}$, GPa | Rms % Error | $E$, GPa | $K$, GPa | $v$ | Number of peaks |
|---|---|---|---|---|---|---|---|
| Sample I-1  | 296.31 | 115.33 | 0.042 | 272.5 | 142.5 | 0.181 | 33 |
| Sample II-1 | 265.98 | 103.78 | 0.029 | 244.9 | 127.6 | 0.180 | 28 |
| Sample II-2 | 266.41 | 104.17 | 0.034 | 245.6 | 127.5 | 0.179 | 33 |
| Sample II-3 | 265.82 | 103.84 | 0.029 | 245.0 | 127.4 | 0.179 | 23 |

It is important to note that quality factor $Q$ for all samples was in the range $(2-6)10^3$ despite the observed difference in elastic modulus (Table 1) and densities. No correlation was found between quality factors and moduli of the samples at low and high frequencies. Presented experimental data for fully dense sample I-1 agree very well with theoretical calculations [16, 18].

Two important comments can be made based on results presented in Table 1. First, there is a consistency between data for samples taken from the same HIPed cylinder (like data for samples II-1 – II-3). This confirms a high level of uniformity of material from the same HIPing run. Second, there is a significant difference between sample I-1 and Samples II-1 – II-3, taken from different HIPing runs. This raises a question of dependence of properties on the sample scale, which needs more detailed investigation. The high accuracy of RUS method makes it probably the most reliable technique to test a quality of material or uniformity of densified material for large scale samples.

We can evaluate expected difference in elastic moduli based on observed difference in densities using theoretical results for material with low density of spherical cavities [23]. Volume fraction of cavities $f$ can be evaluate based on densities:

$$f = (\rho_I - \rho_{II})/\rho_I.$$

Theoretical self-consistent estimate for average value of bulk modulus ($K_{av}$) and Poisson ratio ($v_{av}$) for porous media gives [23]:

$$(K_{av}/K)_{th} = 1 - 3f(1 - v)/2(1 - 2v) + O(f^2);$$

$$(v_{av}/v)_{th} = 1 - f(3 - 1/v) + O(f^2);$$

At $f = 0.04$ and $v = 0.18$ theoretical values are:

$$(K_{av}/K)_{th} = 0.923 \quad \text{and} \quad (v_{av}/v)_{th} = 1.102.$$

Experimental values (taking data for Samples II-1 – II-3 as an average values $K_{av}$ and $v_{av}$ for sample I-1 as K, v) are:

$(K_{av}/K)_{ex} = 0.895$   and   $(v_{av}/v)_{ex} = 0.991$.

Reasonable agreement between theory and experiment allows to make conclusion that observed difference in elastic moduli is due to the porosity of the samples II-1 – II-3.

**Microhardness**

The quality of HIPed material allows to prepare a polished shiny surface of mirror quality suitable for microhardness measurements (Figure 6).

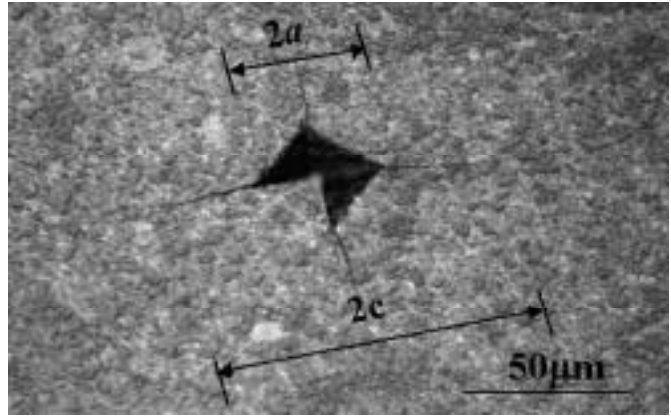

**Figure 6.** Microstructure of the HIPed milled powder and cracks emanating from the indentation corners [7,8].

It is important that the indentations are acceptable for the measurements of Vickers microhardness in brittle materials [24]. The average Vickers microhardness measured under a load of 1 kgf was equal to 10.4 GPa (diagonal length $2a = 41.8$ µm) for the sample prepared from non-milled powder (sample I) and 11.7 GPa (diagonal length $2a = 39.4$ µm) for the sample from milled powder [7,8].

The value of Vickers microhardness equal to 12.79 ± 1.14 GPa at a 4.96-N load was reported for bulk magnesium diboride synthesized using high pressure device (pressure is not specified) from mixture of Mg and B with addition of 2 % of Ta [25]. This microhardness value is close to the value reported in our papers [7,8]. The density of samples was 95 – 97 % of theoretical.

**Fracture Toughness**

Indentation of the HIPed sample presented in Figure 6 demonstrates that it is also acceptable for the evaluation of fracture toughness. As shown in Figure 6, during the microhardness measurements transgranular macrocracks (with average lengths of about 96.6 μm and 86.3 μm calculated from the tip to tip (marked as $2c$) for samples prepared from non-milled and milled powders respectively) emanated from all four corners of the indentation. The presented indentation's shape and cracks are typical for dense ceramic materials [24]. These types of cracks satisfying the condition $c$ $2a$ can be used to estimate the numerical value of fracture toughness using the value of Young's modulus based on the semiemperical equation [26]:

$$K_C = 0.0226\,(EP)^{1/2}\,a\,c^{-3/2},$$

where $E$ is Young's modulus, $P$ is the load used to produce the indentation (9.807 N), and $a$ and $c$ are the lengths of the diagonal and the crack as shown in Figure 6. The Young's modulus of the sample (272.5 GPa) was found based on measured elastic constants using RUS method (Table 1).

The calculated fracture toughness $K_C$ was found to be 2.17 MPa m$^{1/2}$ and 2.41 MPa m$^{1/2}$ for samples HIPed from non-milled and milled powders, respectively. These values of fracture toughness are between the values for single-crystal silicon and single-crystal $Al_2O_3$ (sapphire) [26].

Fracture toughness equal to 4.1 MPa/m$^{1/2}$ at the 147.2-N load was reported in [25], no cracks were detected emanating from the corners of the indent at the load 4.96 N load. Unfortunately no pictures of indents were presented as well no value of elastic modulus was mentioned which authors used to evaluate a fracture toughness. That is why it is difficult to explain the difference between our results and results of [25].

## Machinability

Bulk magnesium diboride has a relatively high microhardness and low fracture toughness. For this reason it is difficult to machine it by turning. High precision samples, for example needed for resonant measurements of elastic properties (group of three parallelepiped samples on left in Figure 7), were prepared from HIPed material in the following way. On the first step rotating diamond saw (60 grid, National Diamond) was used to cut plates from HIPed sample. For a sample diameter of 30 mm it takes about 10 minutes to make a cut. Standard coolant (water plus rust resistance additive were used) on this stage. After this stage plates of magnesium diboride were attached to a metal magnetic plate and sliced into cubes with desirable sizes. Final operation to make sizes parallel and edges sharp was performed using surface grinding machine with 60 grid diamond wheel for the rough removal of surface layer and 400 grid for finishing. Finally a thin deteriorated surface layer on magnesium diboride can be easily removed by dry polishing using fine sand paper producing mirror like surface (Figure 7).

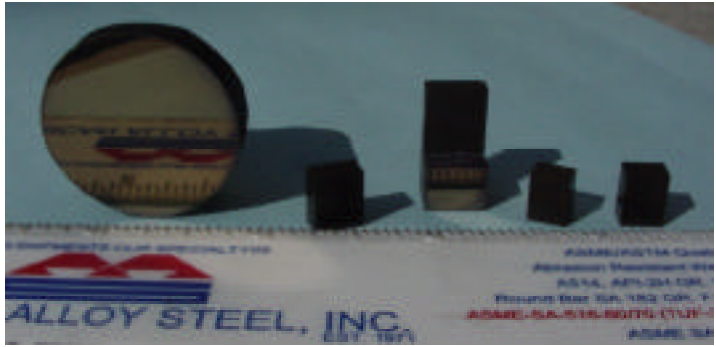

**Figure 7**. Examples of machined samples from HIPed magnesium diboride. Notice mirror like quality of polished sample on left and third from left. Three small parallelepipeds with sharp corners were machined for high accuracy RUS measurements of elastic constants.

## Microstructure of HIPed Material and Superconductivity

Hot isostatically pressed samples have a density which is very close to theoretical density thus improving connectivity of the grains. Additionally to better conductivity of the grains (Figure 8 (a)) the following microstructural features are characteristic for HIPed samples: absence of pores and redistribution of nanosized MgO particles in the bulk of $MgB_2$ grains instead being concentrated at the boundaries ([11], Figure 8 (b)].

TEM brigh-field image (Figure 8(b)) of HIPed sample reveals no porosity or MgO nanoparticles at $MgB_2$ boundaries (which is a typical feature for sintered samples) and high dislocation density. Large strain plastic deformation of $MgB_2$ grains due to pore collapse under the pressure during hot isostatic pressing is responsible for these microstructural features.

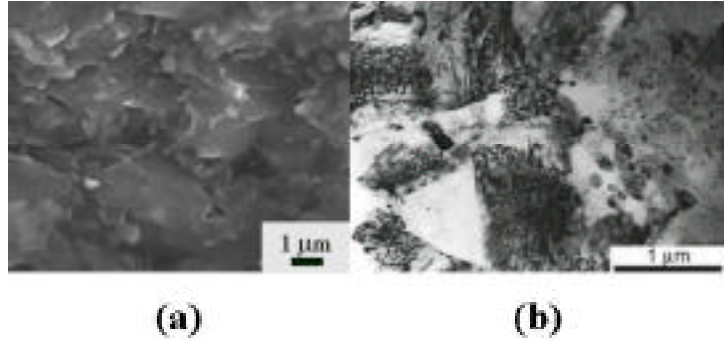

**Figure 8.** SEM micrograph and TEM bright field image of HIPed sample I-1.

Optimization of microstructure, for example, reducing grain size, increasing dislocation density, introducing controlled nanoporosity and nanoparticles are possible ways to improve mechanical and superconducting properties of magnesium diboride.

## Environmental robustness

Superconductivity of porous $MgB_2$ degrades under ambient environment depending on porosity and grain size – superconducting transition becomes broader and Meissner fraction decreases. In contrast, high density HIPed sample remained stable at least for several months under ambient conditions. The details on degradation mechanisms in HIPed samples studied using x-ray photoelecton spectroscopy can be found in [10,12]. The major degradation product from air exposure on the surface of the HIPed sample was $Mg(OH)_x$.

## Superconducting properties

Data on electrical transport, magnetization, and specific heat measurements can be found in [7-9,11,13]. Electrical resistivity of HIPed material at 300K is in the interval 18 – 35 $\mu$ cm, RRR = (300K)/ (40K) in the range 3.1- 3.5.

Superoconducting properties of hot isostatically pressed material most important for practical applications – critical current and its field dependence, upper critical field and irreversibility field are presented in papers [9,13]. These data are depicted in Figures 9 and 10 in comparison with properties of samples sintered at very high pressures [6], which can be considered as bulk samples of very good quality fabricated by alternative technique.

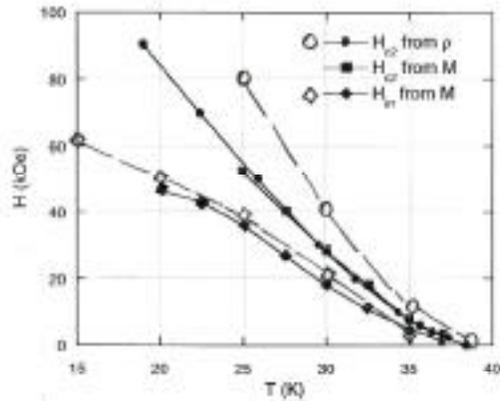

**Figure 9.** Temperature dependence of upper critical field and irreversibility field for HIPed (solid lines) material ([9],[13]) and for material sintered at very high pressure (3.5 GPa, [6]).

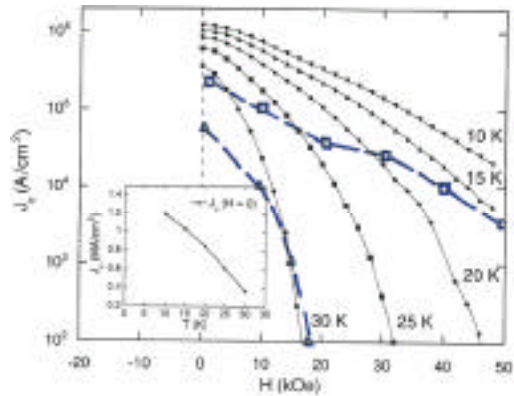

**Figure 10.** Temperature dependence of critical current for HIPed (solid lines) material ([9], [13]) and for material sintered at very high pressure (two broken lines correspond 10K and 30K, [6]).

Both materials have comparable properties with $J_c$ being better at higher fields for HIPed material and $H_{c2}$ (from  ) being higher for high pressure sintered samples. Very important feature of these materials in comparison with sintered samples (for example using PIT method), is that the former do not show a steep drop in critical current at higher fields [6,9,13,11]. This indicates excellent flux pinning in former materials.

Microstructular features of hot isostatically pressed material ([11], Figure 8) are responsible for improving flux pinning. Values of upper critical field for bulk textured magnesium diboride of nearly full density processed by another promising method - hot deformation [27], are below values for HIPed material. Further modification of microstructure are needed to increase density of flux pining centers and superconductting properties behavior of hot isostatically pressed material at higher fields. The promising direction is combination of HIPing with high energy precursor treatment (ball milling, shock loading).

## Microwave Properties of Dense Magnesium Diboride

The fact that magnesium diboride has low electrical resistivity at normal conditions (for HIPed sample as low as 18 $\mu\Omega$·cm [7-9]) makes this material a promising candidate for microwave devices. Machinability of HIPed material and high density ensuring a high quality of surface are attractive qualities for this application also. Microwave study using intracavity method of small samples (presumably of few mm) prepared by high pressure sintering (950 °C, 3 GPa) using commercial $MgB_2$ powder is presented in paper [28]. The density of sample was 2.48 gcm$^{-3}$, resistivity at room temperature was 50 $\mu\Omega$·cm and 20 $\mu\Omega$·cm at 40 K. These values of resistance are substantially higher than for hot isostatically pressed material (18 and 5 $\mu\Omega$·cm correspondingly, [7-9]) and value of density is lower than in HIPed sample (2.666 gcm$^{-3}$). The width of superconducting transition was 0.4 K comparable with the corresponding value for HIPed material [11]. Only normalized value of surface resistivity $R_s/R_{sn}$ at 9.3 GHz can be evaluated in this method because the geometric factor to extract absolute values of $R_s$ should be determined in special experiments with the sample of the same size and of the same effective surface areas, but known conductivity. Normalized surface resistivity $R_s/R_{sn}$ at 20 K was equal 0.05.

Substantial dependence of surface resistance on sample density was reported in [29]. Combination of mechanical and superconducting properties of HIPed materials is obviously attractive for fabrication of microwave devices.

## Summary and Conclusion

The presented analysis demonstrates that hot isostatically pressed magnesium diboride has attractive combinations of properties: high density, environmental robustness, high value of elastic modulus with high level of quality factor $Q$, and fracture toughness comparable to alumina. HIPed magnesium diboride has a reasonable machinability allowing manufacturing of dense, high precision samples of different shapes. Critical current, upper critical field and irreversibility field are among the best reported for this material. Improved field dependence of critical current is due to the specific microstructural features of hot isostatically pressed material (better connectivity of grains, absence of pores and dispersed in the bulk fine MgO particles with sizes around 10-50 nm). Hot isostatic pressing is a very valuable method for basic research aimed toward microstructural modification of $MgB_2$ in combination with other approaches (like mechanical or explosive activation). It is capable for scaling of the sample sizes and for devices with complex shapes.

## Acknowledgements

This work was supported by UCSD start-up fund and by LANL contract 52413-001-02 39. Experiments on hot isostatic pressing and investigation of mechanical properties of magnesium diboride was conducted in cooperation with S.S. Indrakanti and Y. Gu (UCSD). Richard Muzyka (Flow Autoclave Systems, Inc) provided excellent support in installation and extensive use of CIP and HIP machines at UCSD. Investigation of superconducting properties of HIPed samples was conducted by M.B. Maple, N.A. Frederick, B.J. Taylor, S. Li, W.H. Yuhasz (UCSD), investigation of microstructural features, environmental degradation and superconducting characteristics was done by D.E. Peterson, F.M. Mueller, A. Serquis, X.Z. Liao, Y.T. Zhu, J.Y. Coulter, J.Y. Huang, J.O. Willis, N.O. Moreno, J.D. Thompson, R.K. Schulze (LANL). Gary Foreman (UCSD) was very skillful in machining high precision samples of magnesium diboride indispensable in measurements of elastic constant using high precision Resonant Ulrtasound Spectroscopy. Zack Thorington from DRS Inc., was very helpful in introduction of RUS method at UCSD.